\documentclass[sigconf]{acmart}

\usepackage{todonotes}

\newif\ifDEBUG
\DEBUGtrue

\ifDEBUG
    \newcommand{\AH}[1]{\todo[color=cyan,inline]{AH:#1}}
    \newcommand{\AM}[1]{\todo[color=red,inline]{Machiry:#1}}
    \newcommand{\JD}[1]{\todo[color=yellow,inline]{JD:#1}}
    \newcommand{\SA}[1]{\todo[color=green,inline]{SA:#1}}
    \newcommand{\PA}[1]{\todo[color=orange,inline]{PA:#1}}
    
    \newcommand{\KR}[1]{\todo[color=yellow,inline]{Kyle:#1}}
    \newcommand{\LS}[1]{\todo[color=green,inline]{LS:#1}}
    \newcommand{\HP}[1]{\todo[color=green,inline]{HP:#1}}
    \newcommand{\NJE}[1]{\todo[color=red,inline]{NJE:#1}}
    \newcommand{\GKT}[1]{\todo[color=red,inline]{GKT:#1}}
    \newcommand{\KL}[1]{\todo[color=cyan,inline]{KL:#1}}
    \newcommand{\RH}[1]{\todo[color=red,inline]{RH:#1}}
    
\else
    \newcommand{\AH}[1]{}
    \newcommand{\AM}[1]{}
    \newcommand{\JD}[1]{}
    \newcommand{\SA}[1]{}
    \newcommand{\PA}[1]{}
    \newcommand{\KR}[1]{}
    \newcommand{\LS}[1]{}
    \newcommand{\HP}[1]{}
    \newcommand{\NJE}[1]{}
    \newcommand{\GKT}[1]{}
    \newcommand{\KL}[1]{}
    \newcommand{\RH}[1]{}
    
\fi

\usepackage{cleveref}
\usepackage{xspace}

\crefformat{section}{\S#2#1#3}
\crefname{figure}{Figure}{Figures}
\crefname{table}{Table}{Tables}
\crefname{theorem}{Theorem}{Theorems}
\crefname{thm}{Theorem}{Theorems}
\crefname{lemma}{Lemma}{Lemmata}
\crefname{equation}{Eqt.}{Eqts.}
\crefformat{Grammar}{Grammar #1}
\crefname{appendix}{Appendix}{Appendices}
\crefname{listing}{Listing}{Listings}

\newcommand{\eg}{\textit{e.g.,}\xspace}

\usepackage{graphicx} 
\usepackage{wrapfig} 
\usepackage{booktabs} 
\usepackage[frozencache,cachedir=pyg]{minted}
\usepackage{enumitem}
\usepackage{balance}

\AtBeginDocument{%
  }
\setcopyright{acmlicensed}
\copyrightyear{2024}
\acmYear{2024}
\acmDOI{XXXXXXX.XXXXXXX}
\acmConference[TBD]{}{}{}

\begin{document}

\title{Can Large-Language Models Help us Better Understand and Teach the Development of Energy-Efficient Software?}

\author{Ryan Hasler, Konstantin Läufer, \\ George K. Thiruvathukal}
\affiliation{%
  \institution{Loyola University Chicago}
  \city{Chicago, Illinois}
  \country{USA}
}
\email{{rhasler, klaufer, gthiruvathukal}@luc.edu}

\author{Huiyun Peng, Kyle Robinson, Kirsten Davis, Yung-Hsiang Lu, James C. Davis}
\affiliation{%
  \institution{Purdue University}
  \city{West Lafayette, Indiana}
  \country{USA}
}
\email{{peng397, robin489, kad, yhl, davisjam}@purdue.edu}

%
%
%
%
%
%

\renewcommand{\shortauthors}{Hasler et al.}


\begin{abstract}
Computing systems are consuming an increasing and unsustainable fraction of society's energy footprint, notably in data centers.
Meanwhile, energy-efficient software engineering techniques are often absent from undergraduate curricula.
We propose to develop a learning module for energy-efficient software, suitable for incorporation into an undergraduate software engineering class.
There is one major problem with such an endeavor:
  undergraduate curricula have limited space for mastering energy-related systems programming aspects.
To address this problem,
  we propose to leverage the domain expertise afforded by large language models (LLMs).
In our preliminary studies, we observe that
  LLMs can generate energy-efficient variations of basic linear algebra codes tailored to both ARM64 and AMD64 architectures, as well as unit tests and energy measurement harnesses. 
On toy examples suitable for classroom use, this approach reduces energy expenditure by 30–90\%. 
These initial experiences give rise to our vision of LLM-based meta-compilers as a tool for students to transform high-level algorithms into efficient, hardware-specific implementations.
Complementing this tooling, we will incorporate systems thinking concepts into the learning module so that students can reason both locally and globally about the effects of energy optimizations.
\end{abstract}

\maketitle


\pagestyle{plain}



\section{Introduction}
Global climate change poses a serious threat to societal well-being.
Data centers, which account for an estimated 4\% of annual energy consumption in the United States~\cite{epri2024report} and 3\% in the European Union~\cite{eu2024sustainability}, contribute significantly to this issue.
As data center energy consumption is expected to grow, immediate action is needed.
Although advancements have been made in enhancing data center computing system performance along various metrics --- such as reducing latency~\cite{Hua2017}, increasing throughput~\cite{Zhan2012}, and improving parallelism~\cite{Mondal2015} --- knowledge of improving their energy efficiency remains limited~\cite{manotasEmpiricalStudyPractitioners2016}.
\textit{There is little literature on training software engineers to use data center resources with an eye toward energy efficiency}~\cite{mullenLearningDoingHigh2017,qasemGentleIntroductionHeterogeneous2019}.
While energy optimizations have been proposed for individual components~\cite{anagnostopoulouBarelyAliveMemory2012,weiserSchedulingReducedCPU1996,sehgalOptimizingEnergyPerformance2010,sehgal2010evaluating,vasicMakingClusterApplications2009}
and entire systems~\cite{heathLoadBalancingUnbalancing2001,wuImprovingDataCenter2015,townendInvitedPaperImproving2019,chiCooperativelyImprovingData2021},
they have largely focused on hardware.
We have identified a need for improvement in software as well.
Implementing energy-efficient software methods remains challenging, as current approaches rely on heavyweight design approaches~\cite{tebrinkeDesignMethodModular2013,tebrinkeToolsupportedApproachModular2014}, 
pattern catalogs~\cite{pintoComprehensiveStudyEnergy2016, malekiUnderstandingImpactObject2017}, programming languages~\cite{10.1145/3125374.3125382}, or decision frameworks~\cite{manotasSEEDSSoftwareEngineer2014}.

\begin{figure}[t!]
    \centering
    \includegraphics[width=0.90\linewidth]{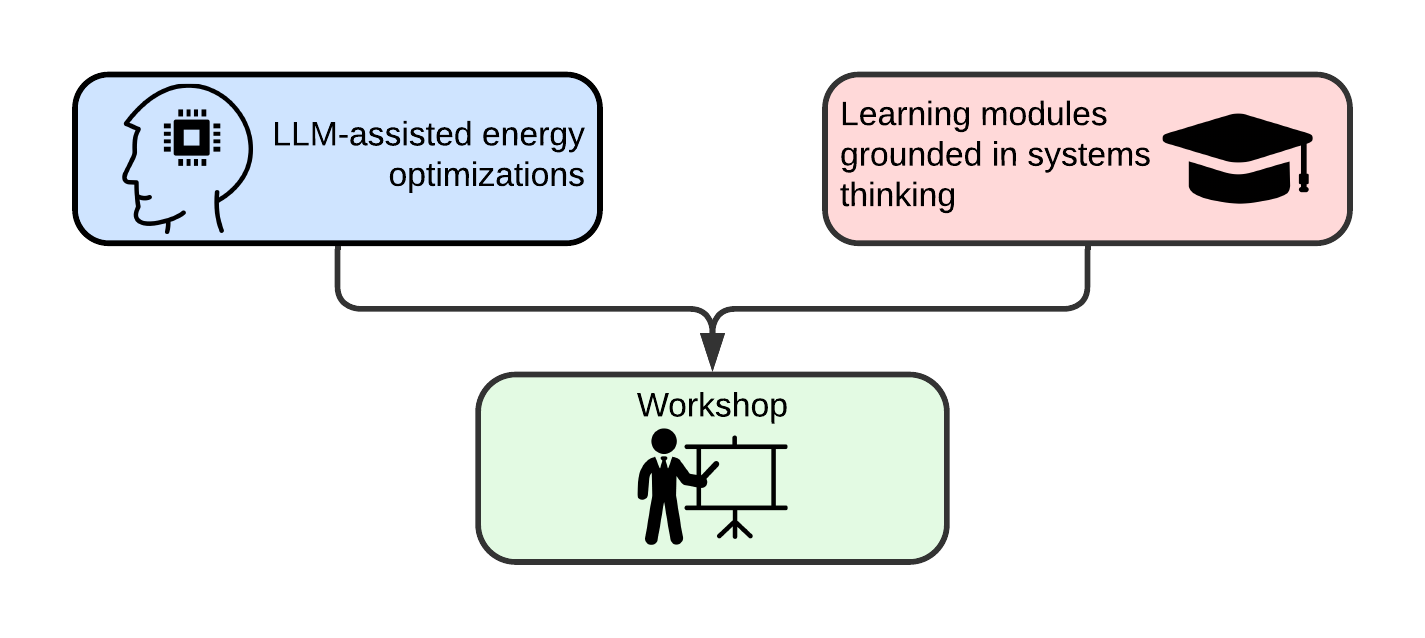}
    \vspace{-0.50cm}
    \caption{
    Research overview:
    we are developing techniques and pedagogy to support
    a two-part learning module. 
    }
    \label{fig:project-flow}
    \vspace{-0.20cm}
\end{figure}

We argue that all software engineers should have access to training on how to write more energy-efficient software. 
They should be able to learn not only what lightweight tools are available to them, but also how systems-level thinking can drive energy-conscious design decisions.
Our preliminary results suggest that large language models (LLMs) might be a useful tool to realize this vision.
We are examining two
educational research questions:

\begin{enumerate}[leftmargin=*, labelindent=0pt]
\item \emph{RQ1:} How can we apply LLMs to help engineers learn to write energy-efficient software?
\item \emph{RQ2:} Is systems thinking an effective approach for design-level reasoning about energy-efficient software engineering designs?
\end{enumerate}




\section{Initial Experiment: GPT-4 as Optimizer}

\paragraph*{Goal}
Our initial objective was to evaluate whether the LLM ChatGPT-v4 (GPT) could assist in developing energy-efficient versions of common algorithms such as the vector dot product. 

\paragraph*{Methodology}

We asked GPT to:
(1) write a basic version of the algorithm in C;
(2) make a parallelized version with OpenMP directives;
(3) use SIMD mode on ARM64 processors;
and finally
(4) leverage both OpenMP and SIMD mode for greater performance and efficiency.

This sequence resembles a homework exercise.
However, note the substantial background knowledge necessary to complete this activity without assistance:
  students would need to know know
    C/C++ (not guaranteed in Python- or Java-first curricula);
    parallel programming libraries such as OpenMP;
  and
    specialized hardware instructions on ARM64.
Such a homework would thus be unsuitable in many undergraduate curricula.
While we do not discourage instructors from offering courses that teach these topics in depth, we believe that any technique that allows less sophisticated students to explore energy-efficient software programming would be valuable to the education community.

LLMs such as GPT provide just this capability.
For the dot product task, GPT proposes a familiar function as a starting point:

\setminted{numbersep=5pt}
\begin{minted}[frame=lines,fontsize=\small,linenos,xleftmargin=9pt]{cpp}
double result = 0.0;
for(int i = 0; i < size; i++) {
    result += a[i] * b[i];
}
\end{minted}

To add OpenMP, GPT inserts a pragma and compilation notes:
\begin{minted}[frame=lines,fontsize=\small,linenos,xleftmargin=9pt]{cpp}
#pragma omp parallel for reduction(+:result)    
\end{minted}

To target the ARM64 Neon SIMD architecture, it leverages the specialized instructions for 4-element float vectors.

\begin{minted}[frame=lines,fontsize=\small,linenos,xleftmargin=9pt]{cpp}
for(int i = 0; i < size; i += 4) {
    a_vec = vld1q_f32(a + i);
    b_vec = vld1q_f32(b + i);
    result_vec = vmlaq_f32(result_vec, a_vec, b_vec);
}    
\end{minted}

\begin{figure}[h!]
    \centering
    \includegraphics[width=0.90\linewidth]{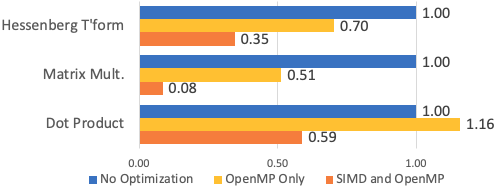}
    \caption{
    \small
      Preliminary results.
      GPT's recommendations improved latency and energy (lower is better) on a Raspberry Pi v4.
    }
    \label{fig:Thrust1-ProofOfConcept}
\end{figure}

\paragraph*{Evaluation}
To evaluate the performance of the algorithms generated by GPT, we measured processing time.
This decision was influenced by the standard model that reducing processing time reduces energy consumption~\cite{4404806,8119702}.
Performance testing was conducted on a Raspberry Pi v4 (ARM) processor. GPT's recommendations resulted in improved latency and energy efficiency (\cref{fig:Thrust1-ProofOfConcept}).
The code and data for the performance tests are available on GitHub at \href{https://github.com/rhasler1/GPTEfficiencyMeasurements}{https://github.com/rhasler1/GPTEfficiencyMeasurements}.

With relatively simple prompts, GPT completed each of these tasks for all three algorithms.
As a proof of our concept, we asked an undergraduate student to conduct this activity --- he did not know OpenMP nor ARM64 instructions.
GPT explained to him everything he needed to know to complete the task.

\section{Future Plans}
Building on these preliminary results, we plan to develop an educational module to improve the training of software engineers to think about sustainability and energy efficiency.
See \cref{tab:LearningModuleOutline} for an outline of the proposed learning module.

\begin{enumerate}[leftmargin=*, labelindent=0pt]
\item \emph{Part One: Energy-Efficient Computing with LLMs:}
We will
  describe energy-proportional computing~\cite{4404806,8119702},
  examine the role of specialized hardware for energy efficiency,
  explain methods for measuring energy usage with industry-grade devices,
  and
  discuss energy benchmarking strategies.
  Additionally, we will demonstrate how LLMs can assist in optimizing algorithms to promote energy efficiency.
\item \emph{Part Two: Systems Thinking as a Framework for Energy Efficiency:}
We will
  introduce students to concepts from Systems Thinking~\cite{arnoldDefinitionSystemsThinking2015}
  and
  apply these principles to energy-efficient software design.
If successful, students will transition from reductionist thinking (\eg considering only local optimizations) to reasoning about larger changes for greater software energy efficiency.
\end{enumerate}

To disseminate our findings to university students and members of the data center programming workforce, we plan to host a regional workshop in partnership with Argonne National Laboratories.
We will also share all materials over the Internet.

\begin{table}[ht]
  \caption{
    Outline of proposed learning module.
  }
  \label{tab:LearningModuleOutline}
  \centering
  \begin{tabular}{cl} 
    \toprule
    {\small \textbf{Unit}} & {\small \textbf{Topic}} \\
    \toprule
    {\small 1} & {\small Data center concepts} \\
    {\small 2} & {\small Energy usage and measure} \\
    {\small 3} & {\small Optimizations (incl. LLM)} \\
    \midrule
    {\small 4} & {\small Systems thinking for energy} \\
    \bottomrule
  \end{tabular}
\end{table}

\section{Conclusion}
The rising energy consumption of software in data centers must not be allowed to contribute to climate change.
We propose to introduce a learning module to improve software engineers' awareness of the energy implications of their code.
We suggest that large language models (LLMs) can help undergraduates integrate the many kinds of knowledge needed to reason about energy.
While our work is still in its early stages, our preliminary results indicate that these approaches have the potential to help students understand when and how to reduce software energy consumption, offering a promising direction for pedagogy.
Given the limited educational literature on enhancing software energy efficiency (especially in data centers), these pedagogical materials could transform how both students and practitioners approach implementing energy-efficient software.
Our approach looks to address some of the immediate energy challenges of data centers, and lay the groundwork for long-term sustainability.

\paragraph*{Acknowledgments}
This work was funded by NSF awards \#2343595 and \#2343596.
This abstract describes part of the research agenda from those awards.

\clearpage

\bibliographystyle{ACM-Reference-Format}
\bibliography{bib/references,bib/PurdueDualityLab,bib/gkt-refs,bib/gkthiruvathukal-cv,bib/laufer-cv,bib/adhocbib}

\end{document}